\documentclass[showpacs, pra,twocolumn,preprintnumbers ,amsmath, amssymb, superscriptaddress, aps]{revtex4-2}
\usepackage[normalem]{ulem}
\usepackage{color}
\usepackage{amsmath,amssymb}
\usepackage{pifont}
\usepackage{amssymb}  
\usepackage{bbold}
\usepackage{float}
\usepackage{subfloat}

\usepackage[caption=false]{subfig}
\usepackage{tikz}
\usepackage{makecell}
\usepackage{subfig}
\usepackage{pifont}   
\usepackage{graphicx} 
\graphicspath{{Figures/}}
\usepackage{dcolumn}  
\usepackage{bm}       
\usepackage{multirow} 
\usepackage{placeins}
\usepackage[colorlinks]{hyperref}
\usepackage{mathtools}
\usepackage{appendix}

\newcommand{\ket}[1]{\left|{#1}\right\rangle}
\newcommand{\bra}[1]{\left\langle{#1}\right|}

\begin{document}
	
	
    \title{Impacts of Intrinsic Noise and Quantum Entanglement on the Geometric and Dynamical Properties of the XXZ Heisenberg Interacting Spin Model}
	\date{\today}
	
	\author{M. Yachi}\email{yachi.m@ucd.ac.ma}
	\affiliation{Laboratory of Innovation in Science, Technology and Modeling, Faculty of Sciences, Chouaïb Doukkali University, 24000 El Jadida, Morocco}
	\author{B. Amghar}\email{amghar.b@ucd.ac.ma}
	\affiliation{Laboratory LPNAMME, Laser Physics Group, Department of Physics, Faculty of Sciences, Chouaïb Doukkali University, El Jadida, Morocco}\affiliation{LPHE-Modeling and Simulation, Faculty of Sciences, Mohammed V University, Rabat, Morocco}
      \author{J. Elfakir}
    \affiliation{LPHE-Modeling and Simulation, Faculty of Sciences, Mohammed V University, Rabat, Morocco}
	\author{M. El Falaki}\affiliation{Laboratory of Innovation in Science, Technology and Modeling, Faculty of Sciences, Chouaïb Doukkali University, 24000 El Jadida, Morocco}\author{S. A. Chelloug}\affiliation{Department of Information Technology, College of Computer and Information Sciences, Princess Nourah bint Abdulrahman University, P.O. Box 84428, Riyadh 11671, Saudi Arabia.}\author{A. A Abd El-Latif}\affiliation{Center of Excellence in Quantum and Intelligent Computing, Prince Sultan University, Riyadh 11586, Saudi Arabia.}\affiliation{Department of Mathematics and Computer Science, Faculty of Science, Menoufia University, Shebin El-Koom 32511, Egypt.}\author{A. Slaoui}\email{abdallah.slaoui@um5s.net.ma}\affiliation{LPHE-Modeling and Simulation, Faculty of Sciences, Mohammed V University, Rabat, Morocco}\affiliation{Center of Excellence in Quantum and Intelligent Computing, Prince Sultan University, Riyadh 11586, Saudi Arabia.}\affiliation{Centre of Physics and Mathematics, CPM, Faculty of Sciences, Mohammed V University in Rabat, Rabat, Morocco.}

	\begin{abstract}
Understanding how intrinsic decoherence affects the interplay between geometry, dynamics, and entanglement in quantum systems is a central challenge in quantum information science. In this work, we develop a unified framework that explores this interplay for a pair of interacting spins governed by an XXZ-type Heisenberg model under an external magnetic field and intrinsic decoherence. We quantify entanglement using the concurrence measure and examine its evolution under decoherence, revealing that intrinsic noise rapidly suppresses entanglement as it increases. We then analyze the Hilbert-Schmidt and Bures distances between quantum states and show how both the degree of entanglement and the noise rate modulate these distances and their associated quantum speeds. Importantly, we demonstrate that the Hilbert–Schmidt speed is more responsive to entanglement and coherence loss than the Bures speed, making it a powerful tool for probing the geometry of quantum dynamics. Moreover, we solve the quantum brachistochrone problem in the presence of intrinsic decoherence, identifying the minimal evolution time and the corresponding optimal entangled states. Finally, we explore the geometric phase accumulated during the system's evolution. Our results show that decoherence hinders geometric phase accumulation, while entanglement counteracts this effect, enhancing phase stability.\par
\vspace{0.25cm}
	\end{abstract}
		
	\maketitle 
	
	\section{Introduction}
\begin{figure}[htbphtbp]
		\centering 
		\includegraphics[width=9cm,height=5cm]{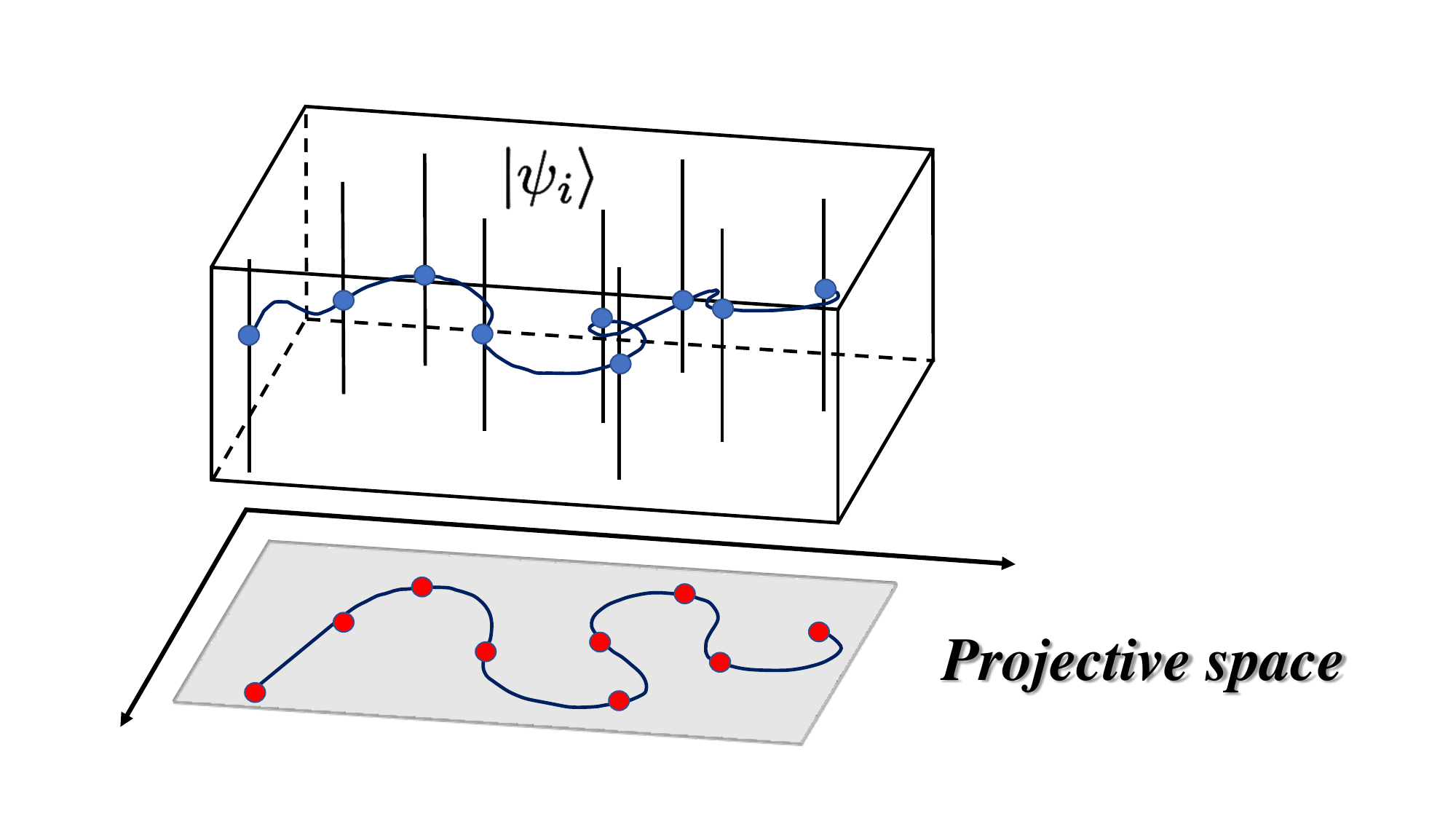}
		\caption{Geometric Representation of Complex Projective Hilbert Space: Mapping Red Points in Projective Space to Corresponding Blue States $\ket{\psi_{i}}$ in Hilbert Space.}
		\label{Dots}
	\end{figure}

Quantum physics operates under principles fundamentally distinct from those of classical physics, presenting phenomena that challenge our intuitive understanding of the macroscopic world \cite{QF}. A striking example is the indistinguishability of quantum states that differ only by a phase shift \cite{Geo3}. This property motivates the mathematical formulation of complex projective Hilbert spaces, $\mathbb{C}P^{n}$, which form the space of physical states for quantum systems. In a given Hilbert space, each point corresponds to a possible quantum state, as illustrated in Fig.\ref{Dots}. These states, represented by vectors $\ket{\psi_{i}}$, encapsulate essential information about the quantum system. Notably, vectors that differ only by a global phase are considered physically equivalent, analogous to how all points on a circle represent the same direction.\par

    Geometric information science is a field that seeks to investigate and elucidate the physical features of quantum systems through a geometric lens, focusing on the manifold of corresponding quantum states. This approach provides a deeper understanding of the intricate relationships and structures underlying quantum phenomena. One of the most prominent features emerging from this geometric perspective is quantum entanglement, which serves as a fundamental physical resource in advanced quantum technologies \cite{Manipu}. Indeed, the degree of entanglement has been linked to the Fubini-Study metric, which defines the geometry of the state space for specific spin systems \cite{XXZ-Fabini}. This relationship has revealed intrinsic connections between the evolution paths of these systems and their entanglement properties. Such connections significantly simplify the approach to solving the quantum Brachistochrone problem, which involves identifying time-optimal trajectories between quantum entangled states. Beyond these dynamic considerations, entanglement has also been investigated through its interplay with the Riemann curvature of the state space, revealing significant insights into the geometric properties underpinning quantum entanglement. Furthermore, in our preceding work \cite{XXZ-Fabini}, we presented a comprehensive discussion of quantum entanglement through the lens of the second Hopf fibration for a two-qubit system. For additional results on the geometric aspects of entanglement, refer to \cite{XXX-Fabini,NZ-Fabini}.\par 

    Another key geometric feature extensively explored in geometric information science is the geometric phase \cite{GeometricPhase1,GeometricPhase2}, which arises from the quantum motion of systems within the associated quantum state space. This geometric property is intimately related to the underlying topology and curvature of the quantum state manifold. It can be obtained by integrating the Berry-Simon connection along the evolution path connecting two different quantum states \cite{Berry1,Berry2,Berry3}. The interconnections between the geometric phase acquired by certain spin systems and quantum entanglement have been well studied in \cite{GeoPhasEntan1,GeoPhasEntan2,GeoPhasEntan3}. Recent studies have highlighted the significant role of the geometric phase in cutting-edge quantum technologies \cite{GeoPhasApp1,GeoPhasApp2}, particularly in its potential applications to implementing unitary gates, which are essential for quantum computing \cite{GeoPhasApp3,GeoPhasApp4}. Furthermore, the experimental implementation of a conditional phase gate has been demonstrated in the realm of nuclear magnetic resonance \cite{GeoPhasApp5,GeoPhasApp6}. For additional findings on the geometric phase, see \cite{GeoPhasApp7,GeoPhasApp8}.\par

In this study, we employ the anisotropic Heisenberg XXZ model, which can be realized in systems such as bosonic atoms \cite{Heisenburg1} or trapped ions \cite{Heisenburg2}. However, understanding Heisenberg interactions becomes even more complex in the presence of intrinsic noise \cite{NoiseXXZ1,Gaidi2024}. The interplay between Hamiltonian dynamics, external magnetic fields, and intrinsic noise creates a highly intricate landscape within the relevant state space. While quantum entanglement offers promising advantages, particularly in the analysis of Bures and Hilbert–Schmidt speeds \cite{Hilbert1,Abouelkhir2023,Hilbert2,Hilbert3,Hilbert4,Hilbert5,Hilbert55}, its practical utility is often limited by the presence of intrinsic noise in open quantum systems \cite{Milburn,OpSys,OpSys1}. Such noise modifies the Schrödinger equation, disrupting coherence in the quantum system \cite{Noise1,Noise2}. In \cite{Milburn}, G. J. Milburn showed that, over sufficiently short time intervals, the evolution of a quantum system deviates from continuous unitary dynamics. Instead, it undergoes a stochastic series of identical unitary operations, with no typical energy loss observed in conventional decay processes, relying solely on phase changes \cite{Hilbert5}. This approach, applicable to a broad range of physically relevant scenarios, holds considerable significance for advancements in quantum information science. Building on previous studies \cite{SpeedBure1,Kuz2,Kuz3,Kuz,Bra1,Bra3} that investigate the geometry of spin systems under noise-free quantum entanglement, we aim to further explore the role of intrinsic noise in shaping quantum dynamics.\par 

The intricate interplay between entanglement, geometry, and decoherence lies at the heart of quantum information processing. Understanding how intrinsic noise shapes this relationship is essential for developing robust quantum technologies. This work aims to unify these aspects through a detailed analysis of a two-spin system evolving under intrinsic decoherence. We examine the system’s evolution under intrinsic decoherence and quantify entanglement using the concurrence measure. Building on this, we derive the Hilbert–Schmidt and Bures distances as functions of entanglement and compare their associated quantum speeds to assess sensitivity to noise and entanglement degree. The study also addresses the quantum Brachistochrone problem by identifying the minimal evolution time and the corresponding optimal entangled states, and concludes with an investigation of the geometric phase, revealing its dependence on both decoherence and quantum entanglement.\par

The present paper is organized as follows: In Sec.\ref{sec2}, we analyze the evolution of two interacting spins governed by the Heisenberg XXZ model under the influence of an external magnetic field and intrinsic noise. In Sec.\ref{sec3}, we quantify the quantum entanglement in the system using concurrence. We then derive the Hilbert-Schmidt and Bures distances in terms of intrinsic noise and concurrence. Next, we establish the Hilbert-Schmidt and Bures speeds and provide a detailed comparative analysis of these speeds with respect to intrinsic noise and quantum entanglement. In Sec.\ref{sec4}, we address the quantum brachistochrone problem in the presence of intrinsic noise by determining the shortest duration required to achieve optimal evolution paths within the noisy quantum state space. In Sec.\ref{sec5}, we investigate the geometric phase acquired by the system during its evolution, focusing on how intrinsic noise and quantum entanglement influence this phase. Finally, we summarize the paper in Sec.\ref{sec6}.
\begin{figure}[H]
		\centering 
		\includegraphics[width=8cm,height=4cm]{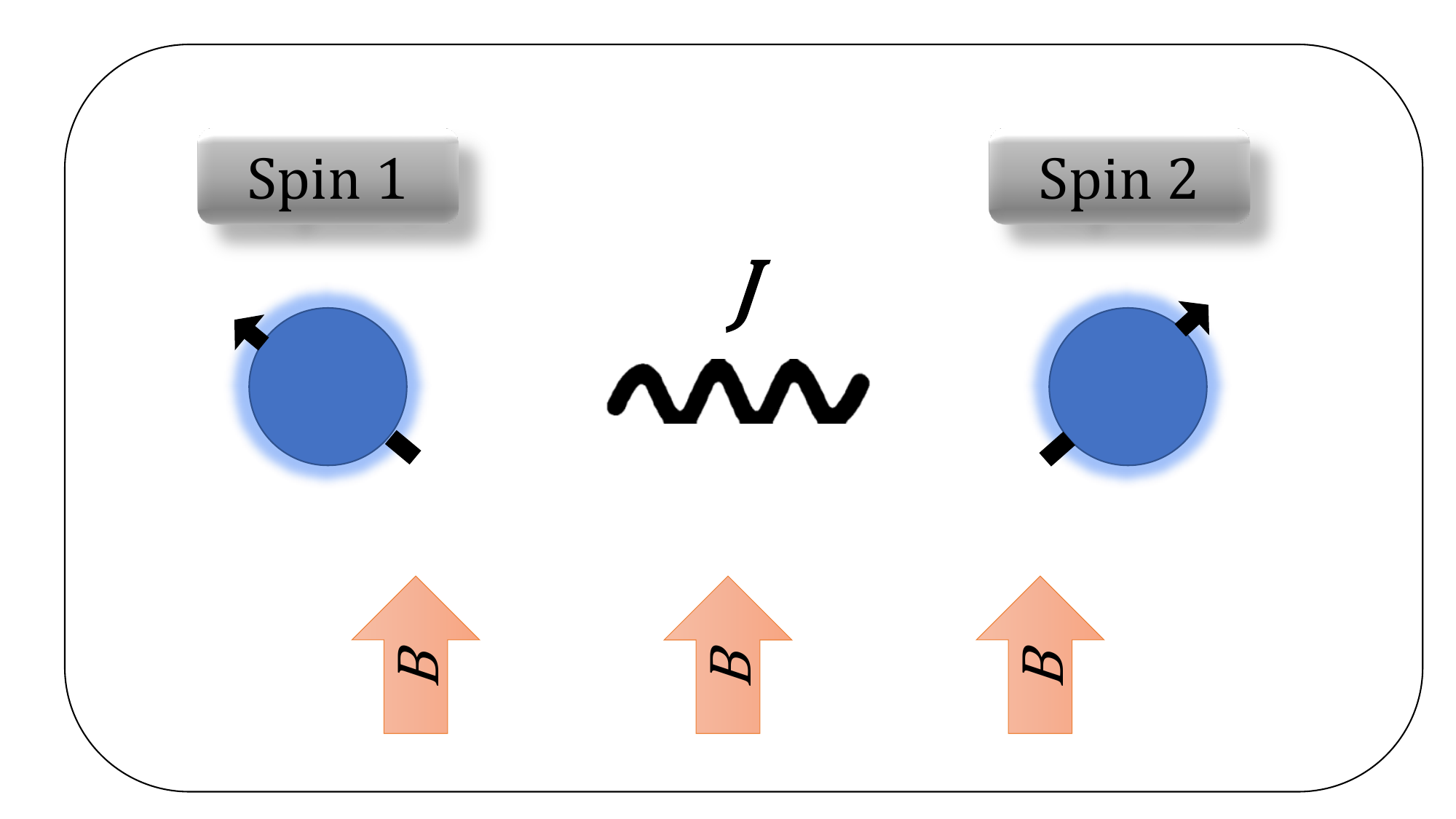}
		\caption{ Illustrative diagram showing the interaction $\mathcal{J}$ between the two spins under the effect of the external magnetic field $B$.}
		\label{2Spins}
\end{figure}
\section{Quantum evolution of two-spin state under intrinsic noise}\label{sec2}

\subsection{Model and eigen-spectrum}
	In this study, we consider a two-spin system described by the Heisenberg XXZ model and subject to an external magnetic field directed along the z-axis, as shown in {\color{black}Fig. \ref{2Spins}}. The system's dynamics is governed by the following Hamiltonian 
	\begin{eqnarray}\label{1}
	\mathcal{H}=\mathcal{H}^{xx}+\mathcal{H}^{z}+\mathcal{H}^{B} 
	\end{eqnarray}
    with 
	\begin{align}\label{2} 
		&\mathcal{H}^{xx}=\mathcal{J}(\sigma^{x}_{1}\otimes\sigma^{x}_{2}+\sigma^{y}_{1}\otimes\sigma^{y}_{2}), \notag\\&  \mathcal{H}^{z}= \gamma\sigma^{z}_{1}\otimes\sigma^{z}_{2},\qquad
		\mathcal{H}^{B}=B(\sigma^{z}_{1}+\sigma^{z}_{2}) 
	\end{align} 
	where $ \sigma^{k}_{i} $ $ (k=x,y,z)  $  are the usual Pauli matrices associated with each spin, $ \mathcal{J} $ stands for the coupling constant between the two spins, $ \gamma $ is a real number characterizing the Heisenberg's anisotropy, and $ B $  denotes the z-component of the external magnetic field applied to the system. In the computational basis $ \mathcal{B} = \{\ket{\uparrow\uparrow},\ket{\uparrow\downarrow},\ket{\downarrow\uparrow},\ket{\downarrow\downarrow}$\}, the Hamiltonian \eqref{1} takes the matrix form 
	\begin{align}
		\mathcal{H}=\begin{pmatrix}
			\gamma+2B& 0 & 0 & 0 \\
			0& -\gamma & 2\mathcal{J} & 0  \\
			0& 2\mathcal{J} & -\gamma & 0  \\
			0& 0 & 0 & \gamma-2B 
		\end{pmatrix}, \label{rho(t)}
	\end{align}
whose eigenstates  are given by 
\begin{align}
		&\ket{\psi_{1}}=\ket{\uparrow\uparrow};\qquad
		\ket{\psi_{2}}=  \frac{1}{\sqrt{2}}\left( \ket{\uparrow\downarrow}+\ket{\downarrow\uparrow}\right);\notag\\&
		\ket{\psi_{3}}=  \frac{1}{\sqrt{2}}\left( \ket{\uparrow\downarrow}-\ket{\downarrow\uparrow}\right);\qquad  \ket{\psi_{4}}=\ket{\downarrow\downarrow}.
		\label{(3)}
\end{align}
    and the corresponding eigenvalues read
	\begin{align}
		&E_{1}= \gamma + 2B  ;\qquad E_{2}= -\gamma + 2\mathcal{J};\notag\\& E_{3}= -\gamma - 2\mathcal{J};\qquad  E_{4}= \gamma - 2B    \label{(2)}.
	\end{align}
Having identified the eigenspectrum of the Hamiltonian, we will proceed to examine the dynamics of the system in the following subsection.
	\subsection{Quantum evolution under intrinsic decoherence}
    To explore the impact of intrinsic noise on the dynamics of the two-spin system, we adopt the Milburn decoherence model, which goes beyond the standard Markovian approximation \cite{OpSys1, Markovian1}. This framework introduces the 
    intrinsic decoherence as a fundamental modification to the system's unitary evolution, accounting for random interruptions in the time evolution at a microscopic level. As a result, the time evolution of the density matrix under intrinsic noise is governed by a modified master equation originally proposed by Milburn \cite{Milburn, Markovian}
	\begin{eqnarray}
		\overset{\textbf{.}}{\mathcal{D}}(t) = -i[\mathcal{H},\mathcal{D}(t)]- \dfrac{1}{2\alpha}[\mathcal{H},[\mathcal{H},\mathcal{D}(t)]],
		\label{(4)}
	\end{eqnarray}
	where we take $\hbar = 1$, which implies that energy is expressed in terms of frequency, $\overset{\textbf{.}}{\mathcal{D}}(t)$ denotes the time derivative of the density matrix $\mathcal{D}(t)$ of the bipartite system, and $\alpha$ represents the intrinsic noise rate. To analyze this evolution, we consider the system initially prepared in the pure state $\mathcal{D}(0) = \ket{\downarrow \uparrow}\bra{\downarrow \uparrow}$. Using this state $ \mathcal{D}(0) $ together with  the equations (\ref{(2)}) and (\ref{(3)}), we obtain the evolved state of the system as follows 
	\begin{align} 
		\mathcal{D}(t)=\begin{pmatrix}
			0& 0 & 0 & 0 \\
			0& u_{22}(t) & u_{23}(t) & 0  \\
			0& u_{23}^{*}(t) & u_{33}(t) & 0  \\
			0& 0 & 0 & 0 
		\end{pmatrix}, \label{D}
	\end{align}
	whose components are given explicitly by
	\begin{eqnarray} 	\label{r23} 
		u_{22}&=& \dfrac{1}{2}\left(1-e^{-4\alpha \mathcal{J} \eta } \cos{2\eta}\right),
        \\ \nonumber
		u_{33}&=& \dfrac{1}{2}\left(1+e^{-4\alpha \mathcal{J} \eta } \cos{2\eta}\right)\\  \nonumber 
		u_{23}&=&-\dfrac{i}{2} e^{-4\alpha \mathcal{J} \eta} \sin{2\eta}. 
	\end{eqnarray}
	here we put $\eta=2\mathcal{J}t$. This simplified form of the two-spin density matrix \eqref{D} significantly facilitates the eigenvalues and their corresponding eigenstates, in particular, in the determination of the geometric phase that can be accumulated during the evolution of the two-spin state. Indeed, the eigenstates of the density matrix \eqref{r23} are found as
    \begin{align}\label{(11)}
        \ket{p_{2,3}}&=e^{3 \alpha \mathcal{J} \eta}\cos{\eta} \left(\ket{\uparrow\downarrow}+\ket{\downarrow\uparrow}\right)\pm\notag\\& i\sin{\eta} \sqrt{\frac{1}{2}(1-\cos{2\eta}+e^{6 \alpha \mathcal{J} \eta}(1+\cos{2\eta}))}\left(\ket{\uparrow\downarrow}+\ket{\downarrow\uparrow}\right);\notag\\& \ket{p_{1}}=\ket{\uparrow\uparrow};\hspace{1cm} \ket{p_{4}}=\ket{\downarrow\downarrow},
    \end{align}
while the corresponding eigenvalues are written 
\begin{align} \label{(11p)}
 &p_{2,3}=\dfrac{1}{2}\left(1\pm e^{-4 \alpha \mathcal{J} \eta}\sqrt{\frac{1}{2}(1-\cos{2\eta}+e^{6\alpha \mathcal{J} \eta}(1+\cos{2\eta}))} \right);\notag\\& p_{1,4}= 0
	\end{align}
In what follows, we consider the dynamical parameter $\eta$ ranging over the interval $[0, 2\pi]$. The following section will be devoted to the quantification of the entanglement degree presented in the system under study. Subsequently, we will analyze how quantum entanglement relates to various distances, their speeds, and the intrinsic noise rate.
	\section{Quantum entanglement, intrinsic noise, and geodesic distances between quantum states \label{sec3}}
	After conducting the dynamics of the two-spin system as described in \eqref{D} under intrinsic decoherence, we will now investigate the Hilbert-Schmidt and Bures distances, as well as the corresponding speeds through the lens of the quantum entanglement and the intrinsic noise rate.
	\subsection{Quantum entanglement in two-spin state}
	The study of the geometry of entangled states constitutes a significant branch of  quantum entanglement theory \cite{QCE}. A fundamental example of the geometrical representation of correlated states in three dimensions is the depiction of a two-spin state with maximally mixed subsystems \cite{QCE1}. Specifically, any two-spin state can be depicted using the Hilbert-Schmidt basis $\{\sigma^{k}\otimes\sigma^{k}\}$, where $\sigma^{0}=I$. In this framework, the correlation matrix $\mathcal{T}$ is defined with elements $t_{ij}=\operatorname{Tr}(\mathcal{D}\sigma^{k}\otimes\sigma^{k})$, for $k=x,y,z$. Various measures have been proposed to quantify the degree of quantum entanglement \cite{Vidal2002,Hill1997,Woo98,Love2007}, among which concurrence measure. It serves as a direct measure of bipartite entanglement for two-qubit systems and is defined as \cite{Woo98}   
	\begin{eqnarray}
		C=\max[0, \lambda_{1}-\lambda_{2}-\lambda_{3}-\lambda_{4}].
	\end{eqnarray}
	It equals $0$ for a separated state and $1$ for a maximally correlated state. The symbols $\lambda_{i}$  are the square roots of the eigenvalues of $ \mathcal{T} $. In our case, the correlation matrix $ \mathcal{T} $ is defined by
	\begin{eqnarray}
		\mathcal{T}(t)=\mathcal{D}(t)(\sigma_{1}^{y}\otimes\sigma_{2}^{y})\mathcal{D}^{*}(t)(\sigma_{1}^{y}\otimes\sigma_{2}^{y}).
	\end{eqnarray}
	After a simple calculation, we find the expression
	of the concurrence as
	\begin{eqnarray}
		C=e^{-4\alpha \mathcal{J} \eta } |\sin{2\eta}|, \label{(15)}
	\end{eqnarray}
       \begin{figure}[H]
		\centering 
	{	\includegraphics[width=8cm,height=6cm]{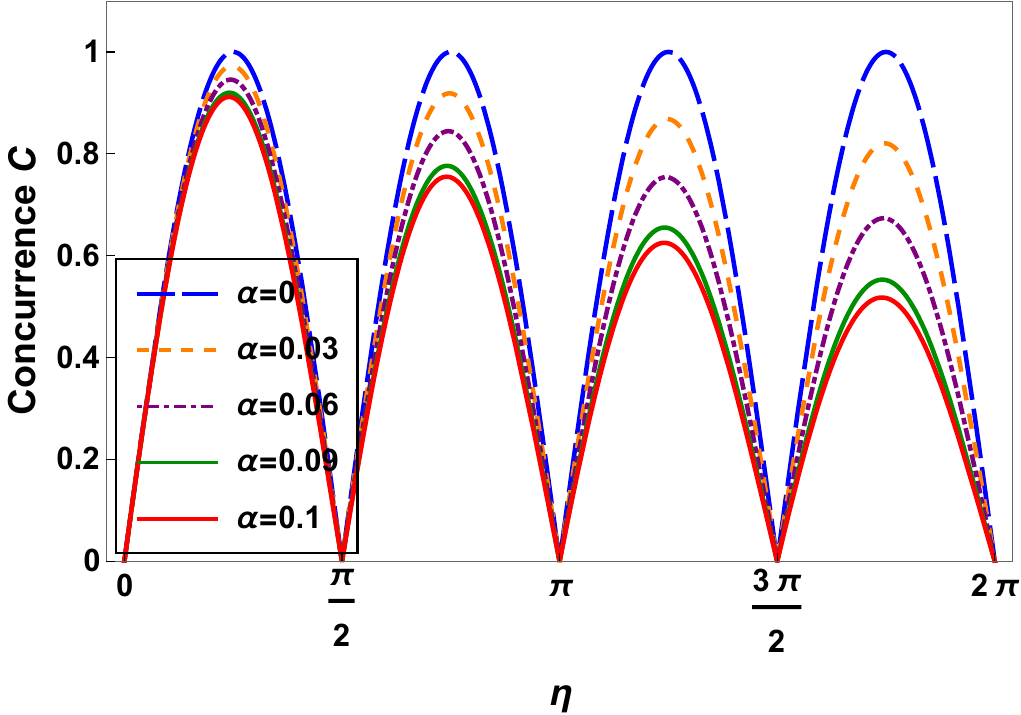}
		\caption{ The behavior of the concurrence \eqref{(15)} versus the dynamical parameter $\eta$, for some intrinsic noise rates by setting  $\mathcal{J}=0.3$.}
		\label{Concurrence}}
	\end{figure}
    Notice that the entanglement is a transient function that oscillates periodically over time. Indeed, it depends on the coupling constant between the two spins and the intrinsic noise rate, while remaining independent of the magnetic field. This independence is attributed to the portion of the Hamiltonian $\mathcal{H}^{B}$ associated with the magnetic field, which exerts a local influence on the two spins, whereas the other components of the Hamiltonian operate on a global scale. In the absence of the intrinsic noise $(\alpha=0)$ the expression of quantum entanglement simplifies to 
	\begin{eqnarray}
		C(\alpha=0)=|\sin{2\eta}|.
	\end{eqnarray}
	This result has already been obtained in our previous paper when we disregarded the effects of intrinsic decoherence \cite{XXZ-Fabini, XXX-Fabini}.\par
 
As shown in Fig. \ref{Concurrence}, we observe that in the noiseless case ($\alpha=0$), the system exhibits periodic and near-maximal entanglement, with well-defined peaks reaching $C \approx 1$, corresponding to the formation of maximally entangled states at regular intervals. As the intrinsic noise rate increases ($\alpha=0.01,0.03,0.06,0.1$), the amplitude of the concurrence oscillations progressively decreases, indicating that decoherence gradually suppresses the system's ability to maintain strong entanglement. At higher noise levels, this suppression becomes more pronounced, with flattened peaks and persistently low minima. This behavior reflects the impact of intrinsic decoherence, which introduces a gradual loss of coherence within the system without invoking an external environment. As a result, the ability of the two spins to sustain or regenerate quantum correlations during their evolution is directly limited by the strength of the intrinsic noise. Overall, the figure demonstrates that the persistence and strength of entanglement are highly sensitive to intrinsic decoherence, and only systems with very low $\alpha$ retain robust entanglement dynamics over time.
	\subsection{From Classical Statistical Distances to Quantum Speeds }
	In classical information theory \cite{Shannon1948,Cover2006}, the divergence between two nearby probability distributions parameterized by $\theta$ is characterized by the Hellinger distance $d(p, q)$ \cite{Jeffreys1946,LeCam1973, Reiss2012}
	\begin{eqnarray}
	 	[d(p,q)]^{2}=\frac{1}{2}\sum_{x}|\sqrt{p_{x}}-\sqrt{q_{x}}|^{2},\label{(1)}
	\end{eqnarray}
   where $p=\left\{p_x\right\}_x$ and $q=\left\{q_x\right\}_x$ denote probability distributions, representing the likelihoods of different outcomes in two distinct experimental settings \cite{Tsybakov2009}. The variable $x$ stands for a discrete random variable. The statistical speed is defined as the rate of change in distance resulting from an infinitesimal variation in the parameter $\theta$
    \begin{eqnarray}
    v(p(\theta_{0}))\equiv\frac{d}{d\theta}d(p(\theta+\theta_{0}),p(\theta_{0})) \label{(2)}.
    \end{eqnarray}
	By extending these classical notions to the quantum setting, and considering two quantum states $ \mathcal{D} $ and $ \mathcal{D}^{'} $, we can write $ p_{x}=\operatorname{Tr}\{E_{x}*\mathcal{D}\}$ and $ q_{x}=\operatorname{Tr}\{E_{x}*\mathcal{D}^{'}\}$ representing the measurement probabilities associated  with the positive operator value measure (POVM) \cite{Luo2004} defined by the set $ {E_{x}\geq0} $ that satisfies $ \sum_{x}E_{x}=\mathbb{I} $. The  quantum distance between these quantum states is 
	\begin{equation}
		L_{B}(\mathcal{D},\mathcal{D}^{'})=\max_{E_{x}}d(p,q)=\sqrt{2-2\sqrt{F(\mathcal{D},\mathcal{D}^{'})}}, \label{(3)}
	\end{equation}
	where $ L_{B}(\mathcal{D},\mathcal{D}^{'}) $ called Bures distance, and $ F(\mathcal{D},\mathcal{D}^{'})\equiv Tr\sqrt{\sqrt{\mathcal{D}}\mathcal{D}^{'}\sqrt{\mathcal{D}}} $ is the fidelity \cite{Jozsa1994}. Thus, the corresponding quantum speed takes the form
	\begin{equation}
	V_{B}(\mathcal{D},\mathcal{D}^{'})\equiv \frac{d}{d\theta}d(\mathcal{D}(\theta+\theta_{0}),\mathcal{D}(\theta_{0}))=\sqrt{\dfrac{F(\mathcal{D}(\theta))}{8}}. \label{(4)}
	\end{equation}
	Further to that, we introduce the concept of Hilbert-Schmidt speed (HSS), a quantum analog of classical statistical speed, based on a quadratic distance measure. Consider the classical distance between two probability distributions $p=\left\{p_x\right\}_x$ and $q=\left\{q_x\right\}_x$, defined by \cite{Bures1}
	\begin{eqnarray}
		[\ell(p,q)]^{2}=\frac{1}{2}\sum_{x}|p_{x}-q_{x}|^{2}. \label{(5)}
	\end{eqnarray}
The corresponding classical statistical speed is given by the rate of change of this distance with respect to the parameter $\theta$ 
 \begin{eqnarray}
	v(p(\theta_{0}))=\frac{d}{d\theta}\ell(p(\theta+\theta_{0}),p(\theta_{0})). \label{(6)}
\end{eqnarray}
Extending this framework to the quantum domain leads to the definition of the Hilbert-Schmidt speed, which quantifies how fast a quantum state evolves geometrically in state space. For a pair of quantum states $\mathcal{D}$ and $\mathcal{D}^{\prime}$, the associated Hilbert-Schmidt distance is defined as \cite{HS1,HS2}
\begin{equation}
	L_{HS}(\mathcal{D},\mathcal{D}^{'})\equiv\max_{E_{x}}\ell(\mathcal{D},\mathcal{D}^{'})=\sqrt{\operatorname{Tr}[(\mathcal{D}^{'}-\mathcal{D})^{2}]}, \label{(7)}
\end{equation}
where the maximization is taken over all POVMs $\left\{E_x\right\}$ that generate the classical distributions $p$ and $q$.
The corresponding Hilbert-Schmidt speed is defined as the instantaneous rate of change of a quantum state with respect to $\theta$, given by \cite{Bures1}
\begin{equation}
	V_{HS}(\mathcal{D})\equiv \max_{E_{x}}v(p(\theta))=\frac{d}{d\theta} L_{HS}(\mathcal{D},\mathcal{D}^{'}). 
    \label{(8)}
\end{equation}
	Having established the fundamental definitions and geometric foundations of the Hilbert-Schmidt and Bures distances, along with their associated quantum speeds, we are now positioned to explore how these structures behave under the influence of intrinsic noise and quantum entanglement. 

    \subsection{Hilbert-Schmidt and Bures distances versus the quantum entanglement and intrinsic noise}
	\subsubsection{Hilbert-Schmidt distance in terms of quantum entanglement and intrinsic noise}
	In simpler terms, the Hilbert-Schmidt distance \cite{HS1,HS2} is a type of distance measure used in quantum information theory.
	For two arbitrary states, $ \mathcal{D}(t) $ and $ \mathcal{D}(t+dt) $, it is given by  
	\begin{eqnarray}
		L_{HS}[\mathcal{D}(t+dt),\mathcal{D}(t)]=\sqrt{\operatorname{Tr}[(\mathcal{D}(t+dt)-\mathcal{D}(t))^{2}]}.
	\end{eqnarray}
This metric is Riemannian, meaning it satisfies the axioms of Riemannian geometry such as smoothness and positive definiteness \cite{Reman}. However, it is not contractive under completely positive maps, meaning that it does not necessarily decrease when two quantum states are transformed by a quantum operation. This characteristic is significant in the study of quantum state transformations, as it suggests that the metric may not accurately reflect the distance between states in all situations involving quantum operations. Explicitly, the Hilbert-Schmidt distance takes the form
	\begin{eqnarray} \label{(18)}
		L_{HS} =2\sqrt{2}e^{-4\alpha\mathcal{J}\eta}
        \sqrt{\mathcal{J}^{2}(4\alpha^{2}\mathcal{J}^{2}+1)}
	\end{eqnarray}
    
    Thus, the Hilbert–Schmidt distance between entangled states is affected by the intrinsic noise rate, the coupling constant, and the system's dynamics. These physical parameters play a key role in shaping the geometric structure of the two-spin system's state space. Moreover, this distance can be directly related to the entanglement level as
	\begin{eqnarray} \label{(19)}
		L_{HS} =\dfrac{2C\sqrt{2}}{C(\alpha=0)}
        \sqrt{\mathcal{J}^{2}(4\alpha^{2}\mathcal{J}^{2}+1)}
	\end{eqnarray}

\begin{figure}[htbphtbp]
		\centering 
		\includegraphics[width=8.6cm,height=5.8cm]{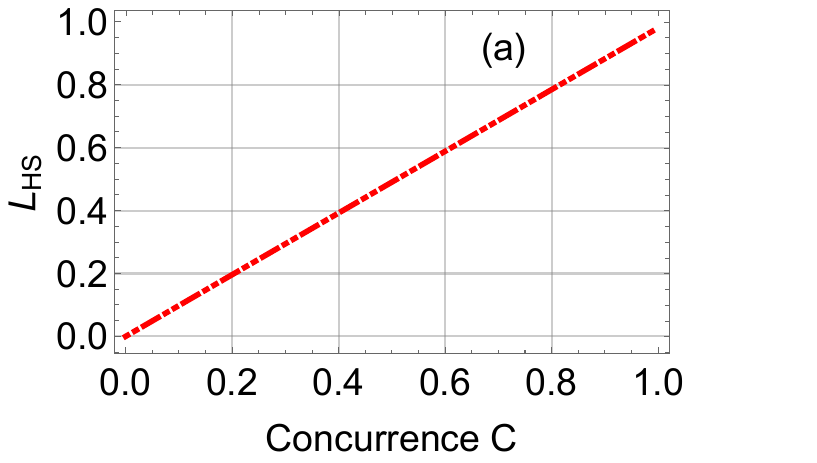}
		\includegraphics[width=8.6cm,height=5.8cm]{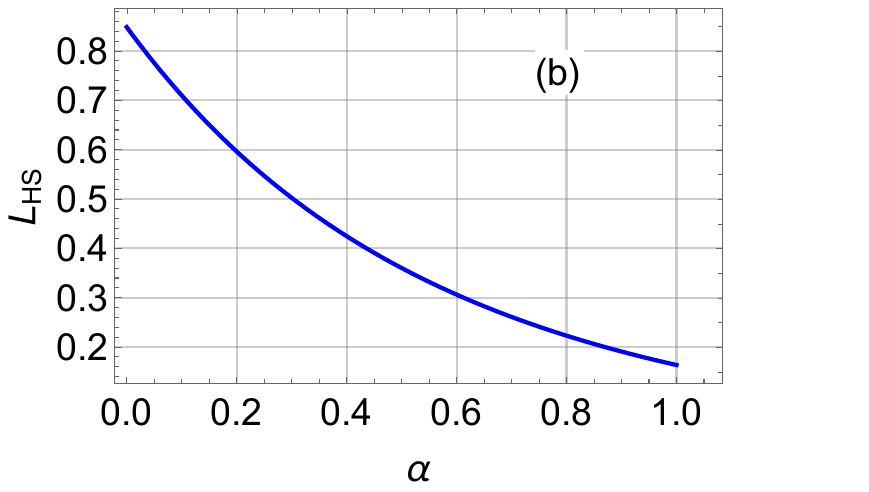}
		\caption{\textbf{(a)} : Hilbert-Schmidt distance \eqref{(19)} versus the concurrence by setting $\mathcal{J}=0.3$, $ \eta=\frac{\pi}{6} $ and $\alpha=0.08$. \textbf{(b)} : Hilbert-Schmidt distance \eqref{(18)} versus the intrinsic noise rate by setting $\mathcal{J}=0.3$ and  $\eta=1.5 $.}
		\label{HSmetric}
	\end{figure}

    This result is important because understanding how concurrence evolves over time allows us to evaluate and control the distance traveled by the system during its quantum evolution \cite{HSmetricpratique1}. This finding is useful for implementing optimal evolution paths, which are essential for designing efficient quantum circuits \cite{QCircuit,QCircuit1}. The behavior of the Hilbert-Schmidt distance with respect to quantum entanglement is clearly illustrated in Fig. \ref{HSmetric}\textbf{(a)}. We notice that the Hilbert-Schmidt distance \eqref{(19)}  increases linearly with the concurrence, indicating a direct proportionality between quantum entanglement and the geometric separation of states. In other words, highly entangled states are geometrically more distant than weakly entangled ones within the quantum state space. From a geometric perspective, increasing entanglement effectively expands the accessible region of the state space, enhancing the distinguishability of quantum configurations. This observation reveals a strong interplay between entanglement and the Hilbert-Schmidt metric: as entanglement grows, so does the geometric "spread" of the state. Therefore, the Hilbert-Schmidt distance can be interpreted as a geometric witness of entanglement, offering a complementary viewpoint to standard entanglement measures \cite{HSmetricpratique2,HSmetricpratique3}.
    
Figure \ref{HSmetric}(\textbf{b}) clearly demonstrates that the Hilbert-Schmidt distance \eqref{(19)} decreases monotonically as the intrinsic noise rate $\alpha$ increases. This behavior indicates that stronger intrinsic noise reduces the distinguishability between quantum states in the system's evolution. Physically, this reflects the decohering effect of noise, which drives quantum states toward classical mixtures, thereby compressing the geometry of the state space. Specifically, the distance between two infinitesimally close states $\mathcal{D}(t)$ and $\mathcal{D}(t+d t)$ becomes progressively smaller with increasing $\alpha$, highlighting a loss of sensitivity to quantum evolution.\par
Furthermore, the metric is more responsive to noise in the low-$\alpha$ regime, where even small increases in $\alpha$ lead to a noticeable reduction in $L_{\mathrm{HS}}$. This suggests that mild decoherence can significantly alter the system's geometric structure, with potential advantages in certain quantum communication scenarios where partial distinguishability of states is beneficial \cite{HSmetricpratique3}. However, in the high-$\alpha$ regime, the distance saturates toward a minimal value, indicating a regime where states become nearly indistinguishable and the capacity for coherent quantum information transfer is severely degraded \cite{HSmetricpratique4}. Thus, this figure emphasizes the dual role of intrinsic noise: while low levels may still preserve useful geometric distinctions, high noise levels collapse the metric structure essential for quantum processing \cite{QCircuit,Dajka2011}.

\subsubsection{Bures distance in terms of quantum entanglement and intrinsic noise}
	The geometric separation between two distinct quantum states can be also quantified using another measure known as the Bures distance \cite{Hilbert55,BuresD,Bures,Bures1,Fid2}. This distance provides a framework for assessing the geometric separation between quantum states, and it is defined as follows
	\begin{eqnarray} \label{(23)}
		L_{B}^{2} = \left(2-2\sqrt{F(\mathcal{D}(t)})\right),
	\end{eqnarray}
	where $	F(\mathcal{D}(t)) $ represents the fidelity of separability, which is expressed in terms of the concurrence as \cite{Fid22}
	\begin{eqnarray} \label{eq21}
		F(\mathcal{D}(t)) =\underset{\sigma \in S}{\max} \mathcal{F}(\mathcal{D}(t),\sigma) = \dfrac{1}{2}\left(1+\sqrt{1-C^{2}}\right),
	\end{eqnarray}
	\begin{figure}[htbphtbp]
		\centering 
		\includegraphics[width=8.5cm,height=5.7cm]{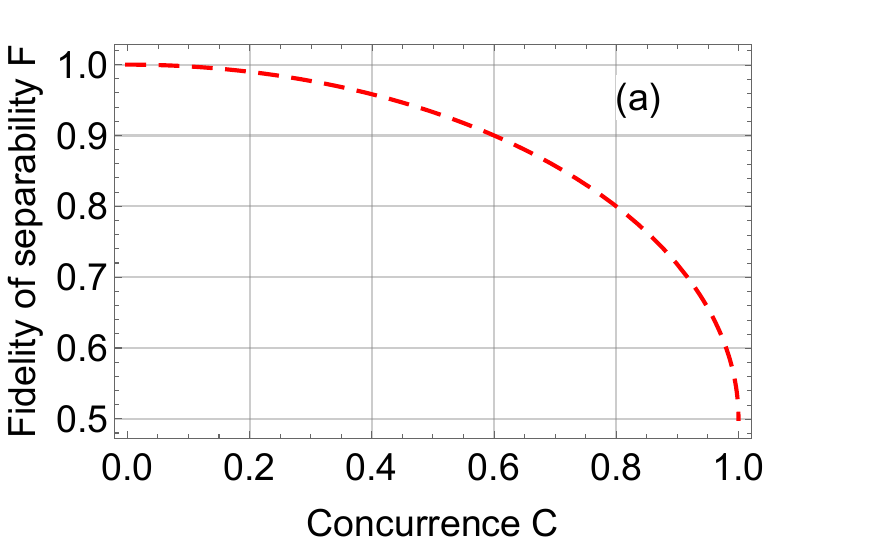}
		\includegraphics[width=8.7cm,height=5.5cm]{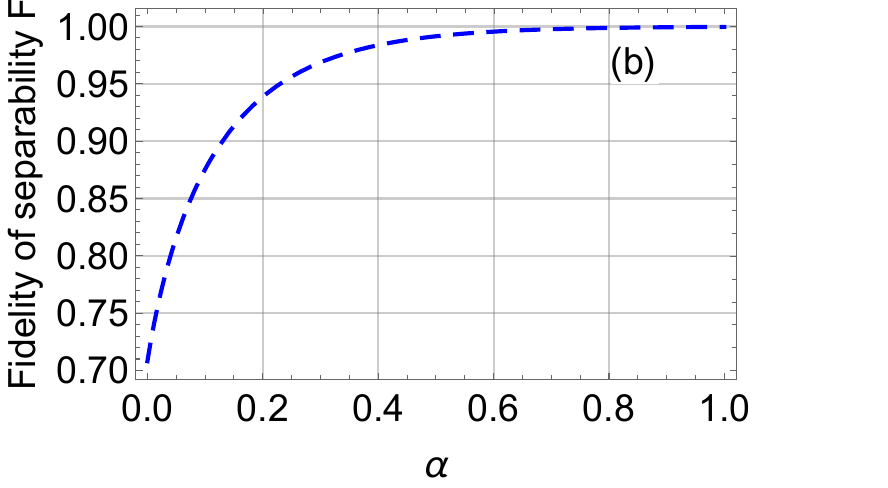}
		\caption{\textbf{(a)} : Fidelity of separability \eqref{eq21} versus the concurrence. \textbf{(b)} : Fidelity of separability \eqref{eq21b} versus intrinsic noise $\alpha$ by setting $\mathcal{J}=0.8$ and  $ \eta=1$.}
		\label{FidelityCon}
	\end{figure}
	where $\mathcal{F}(\mathcal{D}(t), \sigma)$ denotes the Uhlmann fidelity \cite{Fid22}, and $\sigma$ represents an unentangled state within the subspace of separable states of the system $S$. The fidelity is maximized over all unentangled (separable) states $\sigma$ in the set $S$ \cite{Bures1,Fid1,Fid2}.\\
   
    By substituting the result \eqref{(15)} into \eqref{eq21}, the fidelity of separability rewrites
    \begin{eqnarray} \label{eq21b}
		F(\mathcal{D}(t))= \dfrac{1}{2}\left(1+\sqrt{1-(e^{-4\alpha \mathcal{J} \eta } |\sin{2 \eta}|)^{2}}\right).
    \end{eqnarray}
This finding quantifies the closeness of the evolving quantum state to the set of separable states, under the combined influence of spin-spin interaction and decoherence. The exponential term $e^{-4 \alpha \mathcal{J} \eta}$ reflects the dissipative effect of intrinsic noise : as $\alpha, \mathcal{J}$, or $\eta$ increase, the fidelity decreases, indicating a greater departure from separability, and hence a stronger degree of entanglement. The factor $\sin (2 \eta)$ captures the system's intrinsic dynamical oscillations, modulating the impact of noise depending on the phase of evolution. Overall, this result demonstrates that the fidelity of separability is non-trivially shaped by the interplay between unitary dynamics and decoherence, making it a sensitive probe of the residual quantum coherence in the system.\par
 From Fig. \ref{FidelityCon} \textbf{(a)}, we observe that states with low entanglement exhibit a high fidelity of separability, while highly entangled states correspond to significantly lower fidelity values. This inverse relationship highlights the fact that fidelity, quantifying the closeness between quantum states, is strongly affected by the degree of entanglement. In other words, as entanglement increases, a quantum state becomes more distinguishable from any separable state, thereby reducing its fidelity of separability. This behavior underscores a fundamental principle: highly entangled states lie farther from the set of separable states in the quantum state space. Such a distinction is crucial in quantum information tasks, where entanglement plays a key role in enhancing the robustness, nonlocal correlations, and security of quantum communication protocols \cite{FidApp}.\par
    The behavior of the fidelity of separability as a function of the intrinsic noise rate is clearly illustrated in Figure \ref{FidelityCon} \textbf{(b)}. We observe that within the interval $\alpha \in[0,0.6]$, the number of entangled states decreases rapidly, reflecting a strong sensitivity of entanglement to decoherence. Beyond this threshold, i.e., for $\alpha>0.6$, entanglement effectively vanishes, indicating that the two spins become completely disentangled. This transition suggests that at higher levels of intrinsic noise, the quantum coherence required to sustain entanglement is irreversibly lost, and the system evolves toward a fully separable state in which the spins behave independently. Consequently, the generation and preservation of entangled states in such a two-spin system is only viable under low levels of intrinsic noise. This poses a significant challenge for the development of high-performance quantum technologies, where maintaining entanglement in the presence of unavoidable environmental noise remains a major obstacle \cite{Challenge}.\par
\begin{figure}[htbphtbp]
		\centering 
		\includegraphics[width=8.6cm,height=5.8cm]{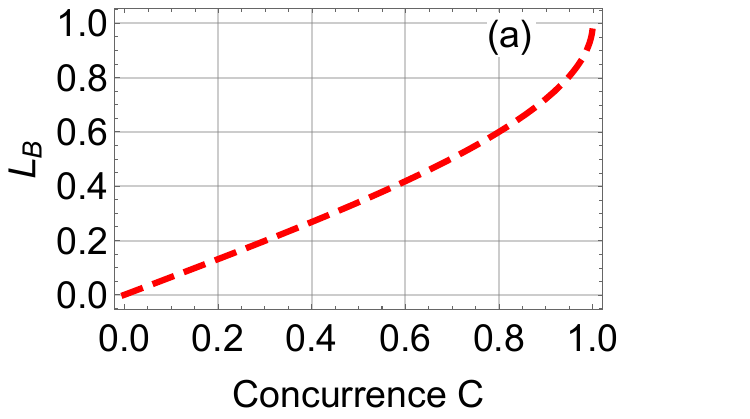}
		\includegraphics[width=8.6cm,height=5.8cm]{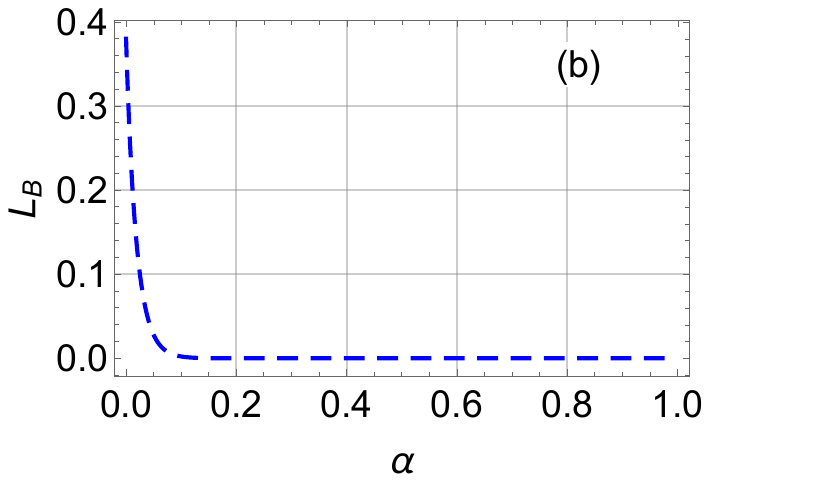}
		\caption{\textbf{(a)} : Bures distance \eqref{eq25} versus the concurrence. \textbf{(b)} : Bures distance \eqref{(26)} versus intrinsic noise $\alpha$ by setting $\mathcal{J}=0.8$ and  $ \eta=16 $.}
		\label{Buremetric}
	\end{figure}

	On the other hand, we explicitly express the Bures distance between the entangled quantum states as
	\begin{eqnarray}
		L_{B} = \dfrac{1}{\sqrt{2-\sqrt{2}}} \left(\sqrt{2-\sqrt{2+2\sqrt{1-C^{2}}}}\right). \label{eq25}
	\end{eqnarray} 
	This result allows us to characterize the geometry of the entangled state manifold based on the degree of quantum entanglement, which serves as a quantifiable physical measure. This approach enables the exploration and experimental measurement of various structures that define the states within the manifold described in equation (\ref{eq25}).  These geometric and dynamic structures encompass the dimension of the state space, the rate of evolution, and the distance between quantum states. As evidenced by equation (\ref{eq25}), the Bures distance is zero for separable states $(C=0)$ and reaches a value of one for maximally correlated states $(C=1)$.  This behavior is illustrated in Fig. \ref{Buremetric} (a). The plot reveals a smooth, monotonically increasing relationship: as the concurrence increases, the Bures distance grows accordingly. This indicates that the Bures metric is sensitive to the degree of quantum entanglement, the states that are highly entangled are located further apart in the Bures geometry. Moreover, the growth is nonlinear and more pronounced for higher values of concurrence, suggesting that the Bures distance responds more strongly to transitions near maximal entanglement than in the weakly entangled regime. This reinforces the geometric interpretation that entanglement stretches the quantum state space, making highly entangled states more distinguishable in a geometric sense. The Bures distance can also be reformulated as a function of the intrinsic noise rate by substituting the eq. (\ref{(15)}) into (\ref{eq25}). This yields the following expression
	\begin{eqnarray}
		L_{B} = \dfrac{1}{\sqrt{2-\sqrt{2}}} \left(2-\sqrt{2+2\sqrt{1-(e^{-4\alpha \mathcal{J} \eta } |\sin{2 \eta}|)^{2}}}\right)^{\frac{1}{2}}. \label{(26)}
	\end{eqnarray}
    This outcome enables us to assess the impact of intrinsic noise on the Bures distance. 
     As illustrated in Fig. \ref{Buremetric} \textbf{(b)}, the Bures distance \eqref{(26)} exhibits a rapid and steep decline in $L_B$ as $\alpha$ increases from zero, followed by a plateau at low values for higher noise rates. This behavior demonstrates that intrinsic decoherence strongly suppresses the Bures distance, particularly in the low-noise regime where coherence and entanglement are still present. Once the noise rate becomes sufficiently large, the system's quantum correlations are almost entirely destroyed, and the Bures distance saturates at minimal values. This suggests a geometric contraction of the accessible quantum state space under decoherence.
	\subsection{Hilbert-Schmidt and Bures speeds versus quantum entanglement and intrinsic noise}
\begin{figure}[htbphtbp]
		\centering 
		\includegraphics[width=8cm,height=6cm]{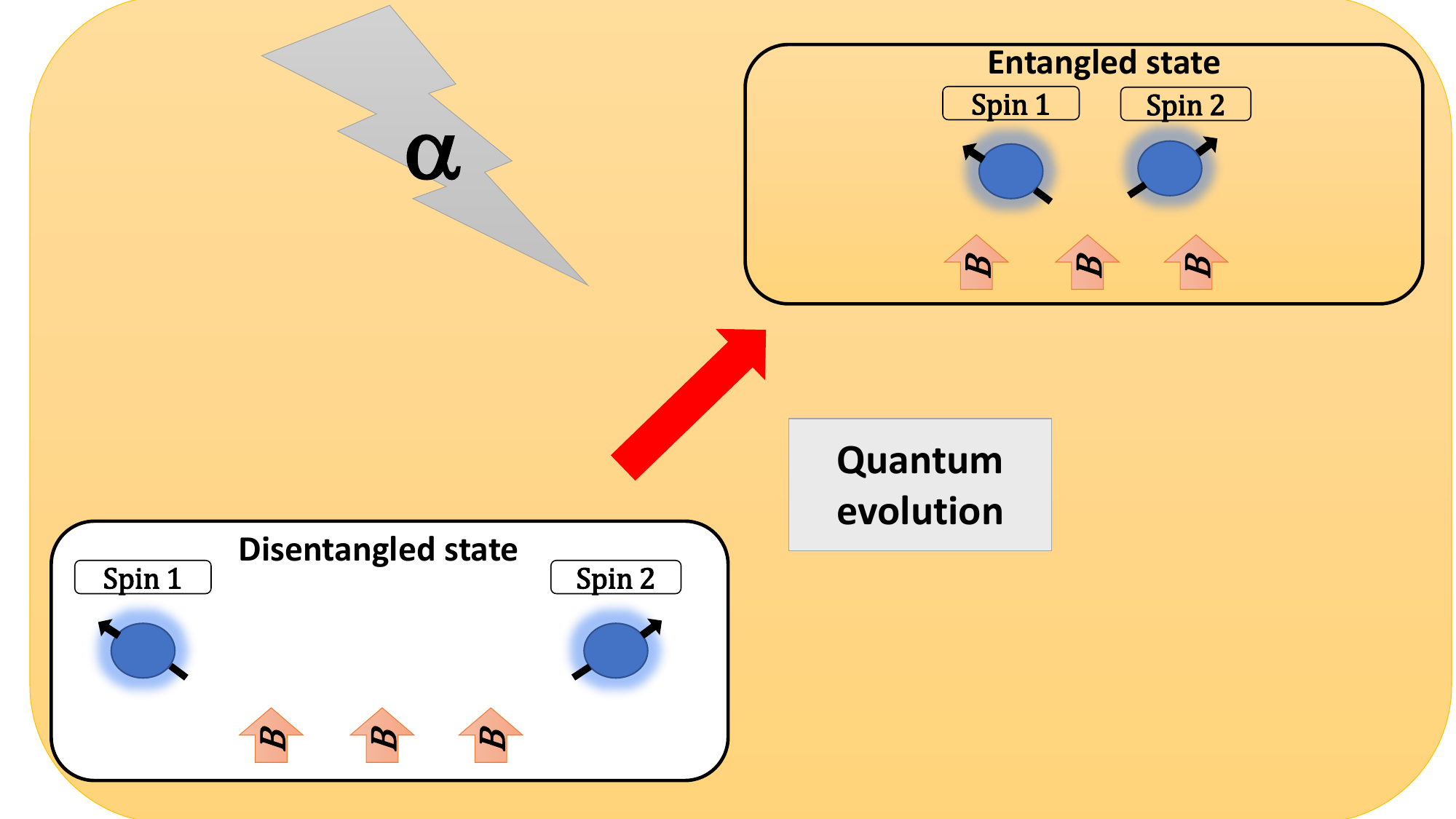}
		\caption{Illustrative Scheme of the quantum evolution from the pure state to the entangled state for the two-spin system under intrinsic decoherence.}
		\label{Dec}
	\end{figure}

	We shall now examine the system dynamics by focusing on quantum entanglement and the effects of intrinsic noise. To achieve this, we begin by calculating the quantum evolution rate, which can be derived using either the Hilbert-Schmidt or Bures distances. This approach enables us to better understand the interplay between entanglement and noise in the evolution of the quantum state (see Fig.\ref{Dec}).
	To evaluate the Hilbert Schmidt speed $ V_{HS} $, we consider a scenario in which the evolution of two spin-1/2 system depends only on time, all other parameters remaining constant \cite{Hilbert2,HSmetricpratique3} :
	\begin{eqnarray} \label{(25)}
		V_{HS}&=&|\dfrac{d}{ d\eta}L_{HS}|. 
	\end{eqnarray}
	Indeed, putting the Eq. (\ref{(18)}) into (\ref{(25)}), the speed $ V_{HS} $ writes
	\begin{eqnarray} \label{(25b)}
	V_{HS}&=&8\sqrt{2}\alpha\mathcal{J}e^{-4\alpha\mathcal{J}\eta}\sqrt{\mathcal{J}^{2}(4\alpha^{2}\mathcal{J}^{2}+1)}.
	\end{eqnarray}
    Thereby, the Hilbert-Schmidt speed between entangled states is affected by intrinsic noise, the coupling constant between spins, and the evolution time. Consequently, these physical parameters play a crucial role in shaping the geometric structure of the state space of the two-spin system. Furthermore, we can relate this distance to the entanglement degree contained in the system as 
	\begin{eqnarray} 
		V_{HS} &=&\dfrac{8\sqrt{2}\alpha\mathcal{J}C}{|\sin{2\eta}|}\sqrt{\mathcal{J}^{2}(4\alpha^{2}\mathcal{J}^{2}+1)}.\label{am22}
        \end{eqnarray}
	Thereby, the Hilbert-Schmidt speed $ V_{HS} $ depends explicitly on the quantum entanglement degree between the two interacting spins. The term $|\sin(2\eta)|$ underlines the temporal modulation, introducing periods of acceleration in the system evolution. The intrinsic noise rate $\alpha$ controls the effect of dissipative interactions, modifying the robustness of quantum entanglement. On the other hand, we can derive  Bures speed $ V_{B} $ using the Bures distance. Indeed, it is given by \cite{Hilbert55,SpeedBure1}
	\begin{eqnarray} \label{eq28}
		V_{B} =\dfrac{d}{ d\eta}L_{B} =\sqrt{\dfrac{F(\mathcal{D}(t))}{8}}.
	\end{eqnarray}
	Employing the equations (\ref{eq21}) and (\ref{eq28}), the Bures speed of the two-spin state takes the form
	\begin{eqnarray} \label{eq29}
		V_{B}= \dfrac{1}{4}\sqrt{1+\sqrt{1-(e^{-4 \alpha \mathcal{J} \eta}|\sin{2\eta}|)^{2}}}.   
	\end{eqnarray}  
    Thus, the Bures speed of the system is also influenced by intrinsic noise, the coupling constant between the spins, and the duration of evolution. As a result, these physical parameters are pivotal in defining the geometric structure of the state space of the two-spin system. Moreover, we can associate this distance with the entanglement level present in the system as follows
	\begin{eqnarray} \label{eq30}
		V_{B}= \dfrac{1}{4}\sqrt{1+\sqrt{1-C^{2}}}.
	\end{eqnarray}  
    This result (\ref{eq30}) illustrates that the Bures speed changes in response to the concurrence. This reveals the sensitivity of the Bures speed to fluctuations in state coherence and levels of entanglement. Furthermore, understanding the evolution of Bures speed in relation to quantum entanglements is crucial for assessing how rapidly information regarding correlated states can be extracted from a quantum system.\par

\begin{figure}[H]
		\centering 
		\includegraphics[width=8cm,height=5.7cm]{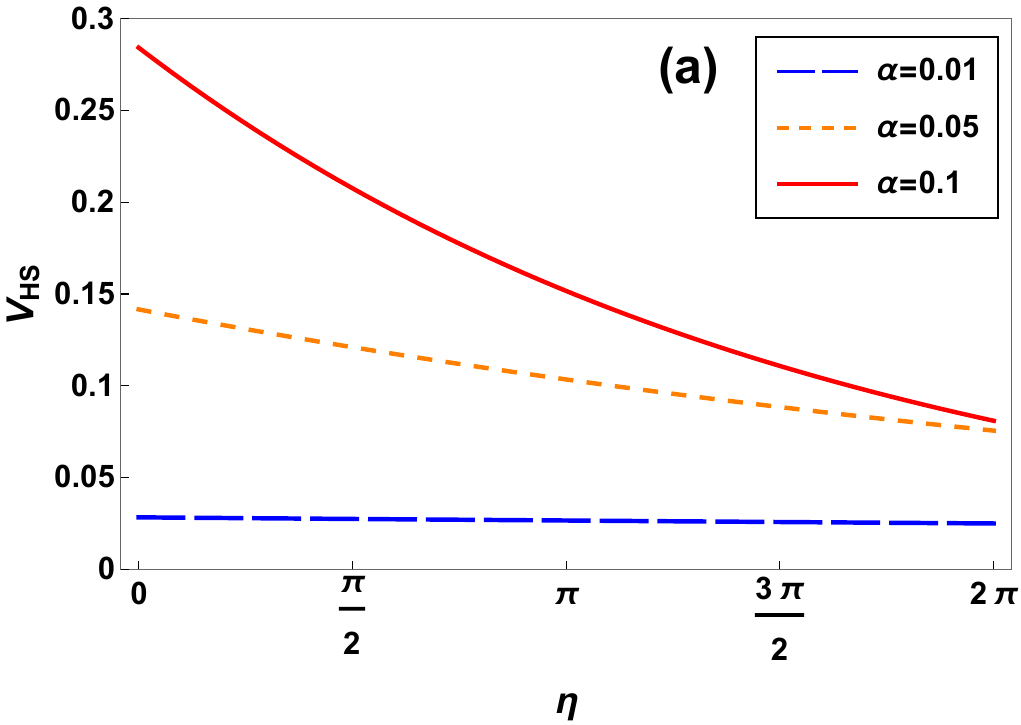} \quad
		\includegraphics[width=8cm,height=5.8cm]{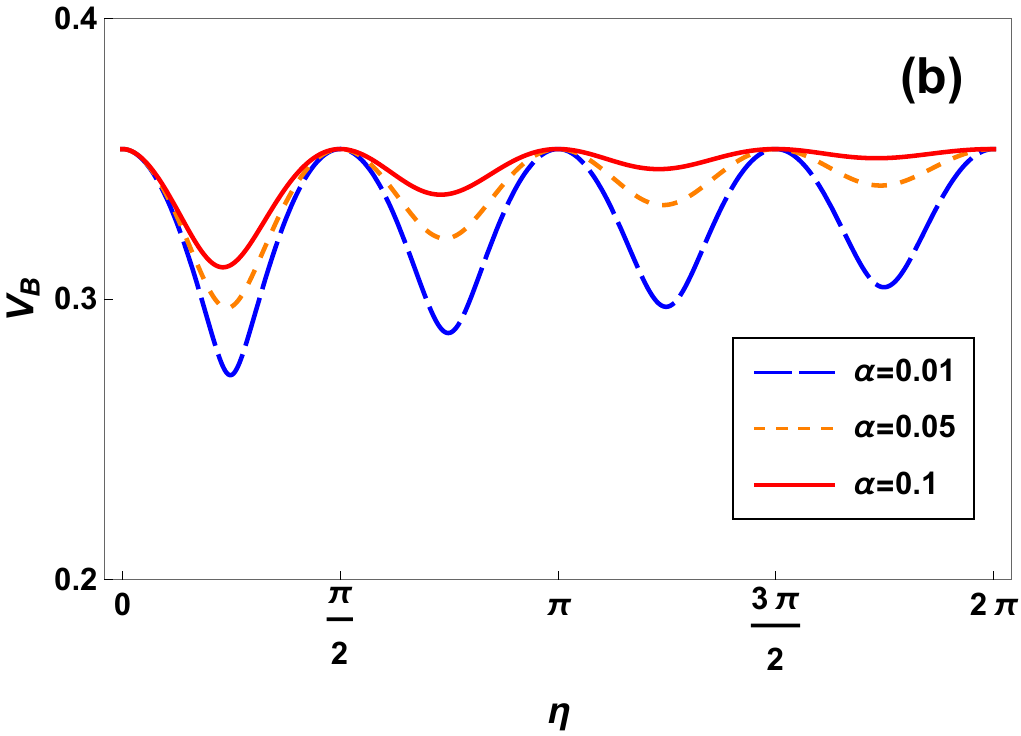}
		\caption{\textbf{(a)} The Hilbert-Schmidt speed \eqref{(25b)} versus $\eta$ for some different intrinsic noise rates,  with $\mathcal{J}=0.5 $. \textbf{(b)} The Bures speed \eqref{eq29} versus $\eta$ for some different intrinsic noise rates,  with $\mathcal{J}=0.5$.}
		\label{Speed}
	\end{figure}

In Fig. \ref{Speed}(a), we present the evolution of the Hilbert-Schmidt speed $V_{HS}$ as a function of the dynamical parameter $\eta$ for various intrinsic noise rates. At a low noise level $\alpha=0.01$, the speed remains nearly constant and very small throughout the evolution, indicating that intrinsic noise has a negligible effect in this regime. However, for higher noise rates $\alpha=0.05$ and $0.1$, $V_{H S}$ starts at a higher value but decreases rapidly with increasing $\eta$. This contrast reveals a two-fold role of intrinsic noise: it initially boosts the evolution speed, yet simultaneously accelerates the degradation of coherence, leading to a faster decay in the system's dynamical behavior. These findings highlight that as the intrinsic noise rate increases, the two-spin system exhibits faster but more transient dynamics. Thereby, we conclude that stronger intrinsic noise more effectively suppresses coherent quantum evolution.

In Fig. \ref{Speed}(b), we depict the variation of the Bures speed $V_B$ as a function of the dynamical parameter $\eta$ for different intrinsic noise rates. At low noise levels ($\alpha=0.01$ and 0.05), $V_B$ displays persistent and well-defined oscillations over time, reflecting robust system dynamics with minimal suppression from decoherence. These oscillations are marked by deeper minima, indicating a high sensitivity of the system to quantum fluctuations in weakly noisy environments. However, as the intrinsic noise rate increases, particularly beyond $\alpha=0.05$, the amplitude of the oscillations begins to attenuate, and the minima of $V_B$ gradually rise. This behavior suggests that stronger intrinsic noise acts as a stabilizing influence: it suppresses abrupt fluctuations in the system's evolution, leading to smoother, more uniform dynamical behavior. Thus, intrinsic noise not only dampens coherence but also contributes to dynamic regularization, enhancing the stability of quantum state evolution under noisy conditions.\par
    
These results reveal a nuanced relationship between intrinsic noise and the dynamical behavior of quantum systems. At low intrinsic noise levels $(\alpha)$, the system exhibits pronounced transient fluctuations, reflecting high sensitivity to quantum coherence. In contrast, higher values of $\alpha$ suppress these fluctuations, leading to greater dynamical stability. This highlights the critical role of accurate intrinsic noise regulation in maintaining the geometric integrity of spin interactions, which is essential for the dependable execution of quantum communication and computing protocols
\cite{BureSpeedApp1,BureSpeedApp2,BureSpeedApp3,BureSpeedApp4,BureSpeedApp5,BureSpeedApp6}. Notably, the Bures speed $V_B$ emerges as a particularly effective metric for characterizing the dynamical response of the system under varying levels of intrinsic decoherence.\par

A comparative analysis of Figures \ref{Speed}(a) and \ref{Speed}(b) highlights the distinct roles played by different quantum speed measures. The Hilbert-Schmidt speed $V_{H S}$ is more sensitive to the loss of coherence and effectively captures the rapid suppression of quantum dynamics as noise increases. Conversely, the Bures speed $V_B$ remains more robust, retaining oscillatory behavior even at moderate to high noise levels, and thus reflects a form of dynamic regularization induced by intrinsic decoherence. These results underscore the complementary nature of these metrics and offer critical insights into the complex interplay between coherence, entanglement, and noise. Such understanding is vital for the design and control of quantum systems in next-generation quantum technologies \cite{Jahromi2020,Genovese2008}.
\begin{figure}[H] \centering	\includegraphics[width=10cm,height=6cm]{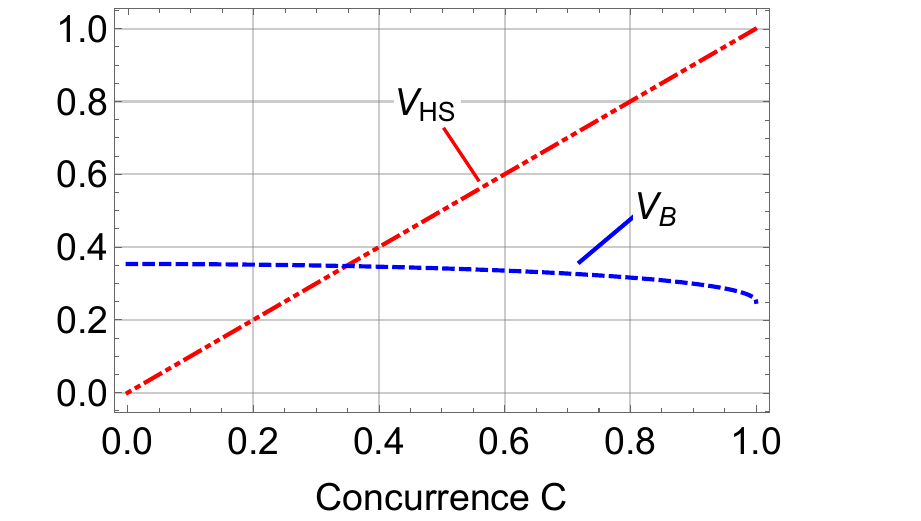}
		\caption{The behavior of Hilbert-Schmidt speed \eqref{am22} and Bures speed \eqref{eq30} versus the concurrence by taking $ \alpha = 0.2$, $ \mathcal{J}=0.65$, and $ \eta=\frac{\pi}{4} $.}
		\label{SpeedCon}
	\end{figure}
Figure \ref{SpeedCon} illustrates the behavior of the Hilbert-Schmidt speed $V_{H S}$ and the Bures speed $V_B$ as functions of concurrence. The Hilbert-Schmidt speed displays a clear linear increase with concurrence, indicating a direct and proportional relationship between the degree of entanglement and the rate of quantum state evolution. This behavior indicates that higher entanglement leads to a faster dynamical response. In the limiting case of a separable (non-entangled) state, the Hilbert-Schmidt speed approaches zero, reflecting the absence of coherent evolution. Therefore, $V_{H S}$ serves as a reliable and sensitive dynamical witness of quantum correlations in the system.\par
In contrast, the behavior of the Bures speed $V_B$ is more nuanced. It exhibits only a slight decrease as concurrence increases, suggesting that it is relatively insensitive to the degree of entanglement. Interestingly, even at low entanglement levels, $V_B$ remains nonzero, indicating that the system can still evolve at a moderate rate despite the absence of quantum entanglement. This observation highlights the role of weak correlations in enabling information flow. As the concurrence grows, the Bures speed decreases slightly, revealing an inverse relationship between entanglement strength and dynamical speed. While this may appear counterintuitive, it can be interpreted as a saturation effect: as strong correlations become established, the need for rapid dynamical exchanges between subsystems diminishes. In this context, the Bures velocity reflects a form of dynamic optimization, where weakly entangled states promote more flexible and faster information transfer, while strongly entangled states stabilize the interaction and reduce dynamic variability.
	\section{Quantum Brachistochrone problem \label{sec4}}  
	Identifying the optimal path through intrinsic noise is essential to achieve the fastest state evolution. However, addressing the quantum brachistochrone problem \cite{Brachi,Brachi1} within the context of the Heisenberg XXZ model under intrinsic noise presents a formidable challenge. This complexity underscores the necessity for innovative approaches to effectively navigate the impact of noise and optimize dynamical processes in quantum systems. The success of this endeavor could pave the way for the development of new quantum control protocols, enabling progress in areas such as quantum error correction, fault-tolerant quantum computing, and the robust processing of quantum information \cite{App}.\par

Now, solving the quantum brachistochrone problem involves determining the minimum time required for the system to evolve from an initial state to a final state within its quantum state space. To achieve this, one seeks to maximize the evolution speed with respect to the concurrence. From Fig. \ref{SpeedCon}, we observe that the Hilbert-Schmidt speed $V_{H S}$ reaches its maximum when the system is maximally entangled, i.e., when $C= 1$, which occurs at $\eta=\frac{\pi}{4}$. From the Eq. \eqref{am22}, the maximal Hilbert-Schmidt speed can be written
\begin{equation}
		V_{HS_{\max}}=V_{HS}(C=1,\eta=\frac{\pi}{4}) = 8\mathcal{J}^{2}\sqrt{2}\alpha\sqrt{4\alpha^{2}\mathcal{J}^{2}+1}.
	\end{equation}
which corresponds to the Hilbert-Schmidt distance
\begin{eqnarray} \label{Lmax}
		L_{HS}(C=1) =2\mathcal{J}\sqrt{2}
        \sqrt{4\alpha^{2}\mathcal{J}^{2}+1}
        \end{eqnarray}

From this, we deduce that the minimum time required to achieve an optimal evolution of the system is
\begin{eqnarray} \label{Lmax}
		t_{\min}= \frac{L_{HS}(C=1)}{V_{HS_{\max}}}=\frac{1}{4\mathcal{J}\alpha}.
        \end{eqnarray}
This result captures the fastest evolution rate achievable under the combined influence of the coupling constant $\mathcal{J}$ and the intrinsic decoherence rate $\alpha$. The time $t_{{\min }}$ represents the shortest duration required for the system to reach a highly correlated (maximally entangled) state. At this critical time $t_{\text {min}}$, the system reaches the optimal entangled state $\mathcal{D}_{\text {op}}^{H S}$, whose explicit form is given by
    {\small
    \begin{align} 
		\mathcal{D}_{op}^{HS}(t_{\min})=\dfrac{1}{2}\begin{pmatrix}
			0& 0 & 0 & 0 \\
			0& \left(1+e^{-2 \mathcal{J} }\cos{\frac{1}{\alpha}} \right) & -ie^{-2 \mathcal{J} }\sin{\frac{1}{\alpha}} & 0  \\
			0& ie^{-2 \mathcal{J} }\sin{\frac{1}{\alpha}} & \left(1-e^{-2 \mathcal{J}}\cos{\frac{1}{\alpha}}\right) & 0  \\
			0& 0 & 0 & 0 
		\end{pmatrix}. \label{(33)}
	\end{align}}
	Consequently, certain off-diagonal elements of the density operator \eqref{(33)} acquire significant values,  indicating the presence of strong coherence and maximal entanglement in the two-spin state. The optimal state $\mathcal{D}_{\text {op }}^{H S}\left(t_{\text {min }}\right)$ is thus the one required to achieve the time-optimal evolution under decoherence. Moreover, this state satisfies the  Milburn evolution equation 
	\begin{eqnarray} 
        \overset{\textbf{.}}{\mathcal{D}}(t_{\min}) = -i[\mathcal{H},\mathcal{D}(t_{\min})]- \dfrac{1}{2\alpha}[\mathcal{H},[\mathcal{H},\mathcal{D}(t_{\min})]],
        \label{am20}
	\end{eqnarray}
	by replacing $t$ with $t_{\min}$ corresponding to the optimal motion of the system.\par Thus, we successfully addressed the quantum Brachistochrone problem by leveraging the system's degree of quantum entanglement, quantified through the Hilbert-Schmidt speed and its associated geometric distance. This approach enabled us to determine the minimal evolution time required for the system to transition optimally between two quantum states. Notably, our analysis highlights that the Hilbert-Schmidt metric is highly sensitive to entanglement and serves as an effective tool for characterizing time-optimal trajectories in the presence of intrinsic decoherence. In contrast, the Bures speed, while geometrically meaningful in other contexts, proves unsuitable for this problem due to its weak dependence on the entanglement degree, rendering it ineffective for identifying entanglement-driven optimal evolutions \cite{Brachi}.\par
    As a natural extension, we explore how this noise-affected evolutionary trajectory leaves a purely geometric imprint on the two-spin system. We therefore turn to an analysis of the geometric phase accumulated along the Milburn‑driven evolution and investigate how it reflects the combined effects of entanglement and intrinsic noise.
	\section{Geometric phase versus the intrinsic noise and quantum entanglement\label{sec5}} 

In this section, we investigate the behavior of the geometric phase acquired by the two-spin system described by the density matrix \eqref{D}, with a particular focus on its dependence on the intrinsic noise rate $\alpha$ and the corresponding degree of quantum entanglement. To ensure mathematical consistency and physical interpretability, we adopt the kinematic approach introduced by Tong et al. \cite{GeometricPhase1,GeometricPhase2}, which is well-suited for computing the geometric phase of mixed quantum states undergoing non-unitary evolution. According to this formalism, the geometric phase accumulated for a cyclic evolution is given by
	\begin{equation} \label{(Phi)}
		\Phi_{g}=\arg \left(\sum_{i}\sqrt{p_{i}(0)p_{i}(\tau)}\langle p_{i}(0)|p_{i}(\tau)\rangle e^{-\int_{0}^{\tau}\langle p_{i}(t)|\overset{.}{p}_{i}(t)\rangle\mathrm{d}t} \right).
	\end{equation} 
    Here, $t \in[0, \tau]$, where $\tau=\eta_{\max } /(2 \mathcal{J})=\pi / \mathcal{J}$ denotes the total evolution time over which the system accumulates the geometric phase. This duration corresponds to a full cycle in the parameter $\eta \in[0, \pi]$, encompassing the complete range of dynamical changes and spin-spin interactions, including the effects of intrinsic noise. At each time $t$, $p_i(t)$ and $\left|p_i(t)\right\rangle$ represent the eigenvalues and corresponding eigenstates of the density matrix $\mathcal{D}(t)$ given in Eq. \eqref{D}. By substituting the analytical expressions for the eigenvalues and eigenvectors from Eqs. \eqref {(11)} and \eqref{(11p)} into the geometric phase formula \eqref{(Phi)}, and using symbolic computation via Mathematica, we obtain the explicit expression for the geometric phase accumulated by the system during its evolution 
	\begin{eqnarray} \label{Phi1}
		\Phi_{g}=\arg \left(\sqrt{\mathcal{A}}\mathcal{B}e^{-(\mathcal{E}+\mathcal{F}+\mathcal{G}+\mathcal{K})}\right),
	\end{eqnarray}
	with
    \begin{align}
        \mathcal{A}&=&\dfrac{1}{2}\left(1+ e^{-4 \alpha \mathcal{J}\eta}\sqrt{\frac{1}{2}(1-\cos2\eta+e^{6 \alpha \mathcal{J}\eta}(1+\cos2\eta))} \right),
    \end{align}

    \begin{widetext}
    \begin{align}
    \mathcal{B}=& i\sin\eta \sqrt{\frac{1}{2}(1-\cos 2\eta+e^{6 \alpha \mathcal{J}\eta}(1+\cos2\eta))}+e^{3 \alpha \mathcal{J}\eta}\cos\eta,
\end{align}
    \begin{align}
        \mathcal{E}&=\dfrac{1}{2}e^{6\alpha \mathcal{J}\eta}\cos^2\eta,\hspace{1cm}\mathcal{G}=\dfrac{1}{16}\left(-4\cos 2\eta+\cos 4\eta+2e^{6\mathcal{J}\alpha \eta}\right)
    \end{align}
    \begin{align}
        \mathcal{F}=&-\dfrac{i}{8}\left[\mathcal{J}\sqrt{1+2\mathcal{J}\eta\left(\frac{\eta}{2\mathcal{J}}+3\alpha\right)}\left(-\frac{4\eta}{\mathcal{J}}-12\alpha+9\mathcal{J}\eta\alpha^{2}\left(1+\frac{9}{2}\eta^{2}\right)+9\mathcal{J}^{2}\alpha^{3}(-13+2\eta^{2})-135\mathcal{J}^{3}\alpha^{4}\eta+1215\mathcal{J}^{4}\alpha^{5}\right)\right.\notag\\&\left.+(1-9\mathcal{J}^{2}\alpha^{2})^{2}(4+45\mathcal{J}^{2}\alpha^{2})\ln\left(\mathcal{J}(\eta+3\mathcal{J}\alpha)+\sqrt{1+2\mathcal{J}\eta\left(\frac{\eta}{2\mathcal{J}}+3\alpha\right)}\right) \right]
    \end{align}
	\begin{eqnarray}\label{(38)} \nonumber
		\mathcal{K}&=&\dfrac{i\sqrt{1+6\mathcal{J}\alpha \eta}}{1701} \left[4+3\mathcal{J}^{2}\alpha\left(-9\alpha+\frac{\eta}{\mathcal{J}}\left(-4+9\mathcal{J}^{2}\alpha\left(3\alpha+\frac{\eta}{\mathcal{J}}\left(2+\mathcal{J}^{2}\alpha\left(81\alpha-2\frac{\eta}{\mathcal{J}}\left(5+3\mathcal{J}^{2}\left(7\frac{\eta}{\mathcal{J}}-27\alpha\right)\alpha\right)\right)\right)\right)\right)\right)\right].
	\end{eqnarray}
    \end{widetext}
	This result (\ref{Phi1}) reveals the complex interplay between the system's dynamics, coupling effects between spins, and intrinsic noise. It results from a subtle combination of dynamical and geometric contributions. The intrinsic noise parameter $ \alpha $ disturbs the coherence properties over time, especially when the system follows a cyclic trajectory in the absence of intrinsic noise. The coupling constant generates complex oscillations, visible in harmonics such as $ \cos{2\eta} $ or $ \cos{4\eta} $, which reflect non-linear effects. The amplitude terms $ \mathcal{A} $ and $ \mathcal{B} $ reflect exponential suppression or amplification dynamics, modulated by intrinsic noise and oscillations dependent on $ \mathcal{J} $ and $ \eta $. Furthermore, $ \mathcal{F} $ contributes with polynomial and logarithmic dependencies reflecting non-trivial phase shifts, while $ \mathcal{G} $ and $ \mathcal{K} $ incorporate higher-order oscillatory and exponential refinements, amplifying intrinsic noise sensitivity.\par
	\begin{figure}[htbphtbp]
		\centering 
	\includegraphics[width=8cm,height=6cm]{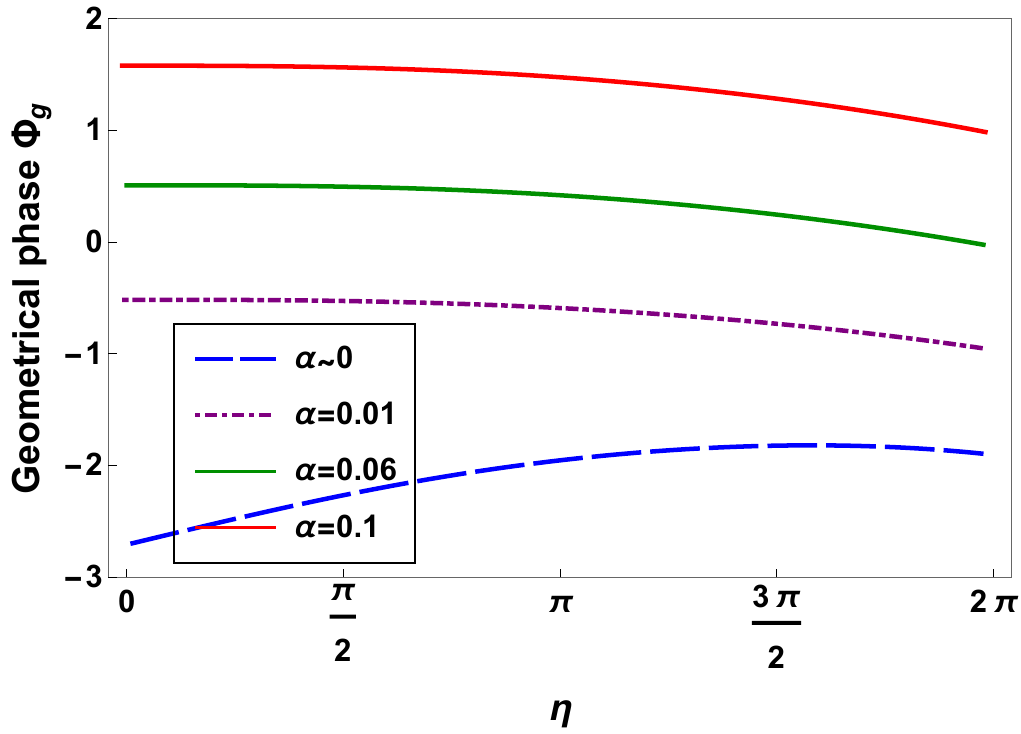}
		\caption{ The geometrical phase \eqref{Phi1} versus the dynamical parameter $\eta $, for some different intrinsic noise rates by setting the coupling constant $\mathcal{J}=0.09$.}
		\label{fig4}
	\end{figure}   

The Fig. \ref{fig4} depicts the variation of the geometric phase \eqref{Phi1} as a function of the dynamical parameter $\eta$ for various intrinsic noise rates. In the absence of noise ($\alpha=0$), the geometric phase increases continuously with $\eta$, reflecting a coherent unitary evolution where quantum interference effects accumulate fully. As the intrinsic noise rate increases $(\alpha=0.01,0.06,0.1)$, the curve flattens progressively and eventually reverses its trend: the phase becomes less negative, then positive, and decreases slowly. This behavior demonstrates that decoherence suppresses geometric phase accumulation by disrupting the system's coherent evolution. Remarkably, even a small amount of intrinsic noise significantly reduces $\Phi_g$, highlighting the high sensitivity of this quantity to environmental perturbations. At higher noise levels, the dynamics become dominated by dissipation, leading to a near-stabilization of the geometric phase. Overall, this figure emphasizes the critical role of intrinsic noise in shaping the geometric properties of the system and underlines the importance of preserving quantum coherence to harness geometric phase in quantum control and information processing protocols \cite{Dasgupta2007,Filipp2009}. \par
Based on the preceding results, it can be inferred that the geometric phase $\Phi_g$ is closely linked to the degree of quantum entanglement within the system. In regimes where decoherence is minimal, $\Phi_g$ accumulates significantly, typically coinciding with strong entanglement between the two spins. Conversely, as intrinsic noise increases, thereby reducing coherence and, consequently, entanglement, the geometric phase tends to diminish or stabilize. Although this relationship is not explicitly depicted in the plots, it is reasonable to anticipate that higher levels of entanglement support a richer and more dynamic accumulation of geometric phase, whereas weakly entangled states, often accompanied by stronger dissipation, result in a more suppressed geometric phase. These insights suggest that, under certain conditions, the geometric phase could serve as an indirect indicator of the system's quantum entanglement \cite{Dajka2011b}.
	\section{Summary  \label{sec6}}
To summarize, we investigated a physical system consisting of two interacting spins described by the XXZ-type Heisenberg model within a magnetic field aligned along the $z$-axis. We examined the system's dynamics under intrinsic decoherence and quantified the quantum entanglement between the two spins using the concurrence measure. Our results highlighted the destructive impact of intrinsic noise, which gradually suppresses entanglement as the noise rate increases. \par
We further analyzed the system's geometry by computing both the Hilbert-Schmidt and Bures distances between quantum states. These geometric distances were shown to depend strongly on the level of entanglement and the decoherence rate. Specifically, a higher degree of entanglement increases the separation between quantum states in the state space, while increased intrinsic noise compresses this space, reducing the distances. In parallel, we showed that weakly entangled states exhibit high separability fidelity, and beyond a critical threshold of intrinsic noise, the system becomes effectively disentangled.\par
In addition, we investigated the Hilbert-Schmidt and Bures quantum speeds, which reflect how fast the quantum states evolve. We demonstrated that the Hilbert-Schmidt speed is more sensitive to both entanglement and intrinsic noise than the Bures speed. Using this insight, we addressed the quantum Brachistochrone problem by reparametrizing the evolution to identify time-optimal paths in the state space. We found that the optimal evolution is achieved when the system is maximally entangled and derived the corresponding minimal time using the Hilbert-Schmidt speed and distance. We explicitly obtained the optimal state by solving the Milburn evolution equation \eqref{am20}, confirming its consistency with intrinsic decoherence dynamics.\par
Finally, we studied the behavior of the geometric phase acquired by the system under decoherence. Our results revealed that intrinsic noise diminishes the geometric phase by disrupting the coherent evolution, while quantum entanglement counteracts this effect by enhancing the phase accumulation. These findings underline the dual role of coherence and entanglement in shaping geometric and dynamical properties of quantum systems.\\

{\bf Acknowledgments:} We are grateful to the Princess Nourah bint Abdulrahman University Researchers Supporting Project Number (PNURSP2025R239), Princess Nourah bint Abdulrahman University, Riyadh, Saudi Arabia. This paper also derived from a research grant funded by the Research, Development, and Innovation Authority (RDIA) - Kingdom of Saudi Arabia - with grant number (13325-psu-2023-PSNU-R-3-1-EF). Additionally, the authors would like to thank Prince Sultan University for their support.\par

{\bf Disclosures:} The authors declare that they have no known competing financial interests or personal relationships that could have appeared to influence the work reported in this paper.\par 

{\bf Data availability:} The datasets used and/or analysed during the current study available from the corresponding author on reasonable request.

\end{document}